\documentstyle[aps,epsf,here,12pt]{revtex}

\tighten
\newcommand{\PostScript}[7]{
\begin{figure}[H]
\begin{center}
\leavevmode
\epsfysize=#1cm
\vspace{#2cm}
\epsfbox{#3}
\par
\parbox{#5cm}{
\vspace{#4cm}
\caption[figure]{\renewcommand{\baselinestretch}{1} \small \normalsize #6}
\label{#7}}
\end{center}
\end{figure}
}

\begin{document}

\newcommand\beq{\begin{equation}}
\newcommand\eeq{\end{equation}}
\newcommand\bea{\begin{eqnarray}}
\newcommand\eea{\end{eqnarray}}

\title{The effect of interactions on Bose-Einstein condensation in a 
quasi two-dimensional harmonic trap}
\author{R. K. Bhaduri$^1$, S. M. Reimann$^2$, S. Viefers$^{2,3}$,
A. Ghose Choudhury$^{4,1}$ and~M.~K.~Srivastava$^5$}
\address
{$^1$Department of Physics and Astronomy, McMaster University,
Hamilton L8S 4M1, Canada\\
$^2$Department of Physics,
University of Jyv\"askyl\"a, P.O. Box 35, 
40351 Jyv\"askyl\"a, Finland\\ 
$^3$NORDITA, Blegdamsvej 17, DK-2100 Copenhagen, Denmark\\
$^4$Department of Physics, Surendra Nath College, Calcutta 700 009, India\\
$^5$Department of Physics, University of Roorkee, Roorkee 
247667, India 
\date{\today}}

\maketitle
\begin{abstract}
A dilute bose gas in a quasi two-dimensional harmonic trap and interacting 
with a repulsive two-body zero-range potential of fixed coupling 
constant is considered. Using the Thomas-Fermi method, it is shown to remain 
in the same uncondensed phase as the temperature is lowered. 
Its density profile and energy  are identical to that of an {\it ideal}
gas obeying the fractional exclusion statistics of Haldane.
\end{abstract}


\narrowtext
\section{Introduction}

The quantum and thermodynamic properties of atom traps with 
reduced dimensionality have been studied by several 
groups \cite{Bag91,Kett96,Gross96,Haug98}. In a recent review 
Dalfovo~{\it et al.}~\cite{Dal99} note that the phase transition 
of bosons in quasi two-dimensional atom traps is an important 
issue for future investigations. In this paper, we consider 
bosons which are trapped in a two-dimensional harmonic oscillator 
and interact by a repulsive zero-range pair potential of a fixed strength. 
Using the finite-temperature Thomas-Fermi (TF) method, we find that 
{\it there is no strict 
phase-transition for such a system, no matter how weak the repulsion}. As the 
temperature is lowered, the chemical potential rises 
smoothly, and only at zero temperature does it match the lowest energy level 
of the trap. By analyzing the equation of state of the 
interacting bose system in the complex fugacity plane, we show 
that the system remains in a single phase at all (real) temperatures.   
We demonstrate that the area integral of the interaction potential in 
two dimensions is related to the statistical parameter of Haldane's  
fractional exclusion statistics (FES)~\cite{Haldane91}. In particular, 
the thermodynamic properties of the bose system interacting with a zero-range 
repulsive potential are found to be of the same form as that of an ideal 
FES gas. 

Consider bosons in an oblate three-dimensional trap with the oscillator 
frequencies $\omega_x=\omega_y=\omega \ll \omega_z$. For a dilute Bose 
gas at low temperatures we take the usual three-dimensional delta function 
pseudo-potential\cite{Fermi36,Huang89} with strength $(4\pi\hbar^2 a/m)$, 
where $a$ is the 
s-wave scattering length, and which in first-order  
Born-approximation yields the correct low-energy quantum scattering result.    
For low enough temperatures, and assuming that the 
expectation value of the pseudo-potential is much smaller than 
$\hbar\omega_z$, we may restrict the Hilbert space 
in the $z$-direction by setting the oscillator 
quantum number $n_z=0$. The quantum dynamics is then 
determined by an effectively two-dimensional Hamiltonian, given by 
\beq
H=\sum_{i=1}^N \left({p_i^2\over {2m}}+{1\over 2}m\omega^2 r_i^2\right)+
{2\pi\hbar^2\over m} g \sum_{i<j}^N \delta({\bf r}_i- {\bf r}_j)~,
\label{tr1}
\eeq
where the momenta and coordinates are planar vectors. 
Such a Hamiltonian has 
been studied recently by several groups~\cite{Haug98,Pit97,Ber98} and also by 
Mullin~\cite{Mull97b,Mull98a}. The dimensionless coupling constant $g$ in the 
two-dimensional delta function interaction may be related to the scattering 
length $a$ for $s$-wave scattering in three dimensions. 
This is done by taking the expectation value of the three-dimensional 
delta function with respect to the  
harmonic oscillator one-dimensional ground state wave function with $n_z=0$. 
We then obtain the effective two-dimensional interaction given in 
Eq.~(\ref{tr1}), with the dimensionless coupling constant   
\beq
g=\sqrt{2\over \pi}\left({a\over b_z}\right)~.
\label{small}
\eeq  
Here $b_z=\sqrt{\hbar/m \omega_z}$ is the length scale of confinement in the 
$z$-direction. 
This quasi two-dimensional description of the actual 
three-dimensional trap is valid only for a dilute system 
in which the scattering length $a \ll b_z \ll b $~\cite{Ber98}, where 
$b=\sqrt{\hbar /{m\omega}}$ is the oscillator length scale in the 
$x$-$y$~plane. It follows 
from Eq.~(\ref{small}), therefore, that $g\ll1$ in Eq.~(\ref{tr1}). 
For an estimate of the parameter $g$, take $b= 3$
micron, which corresponds to $\hbar\omega \simeq 0.6$ nK. 
An {\it ideal} gas in a two-dimensional harmonic trap would have  BEC at 
the critical temperature $T_c^0=(6/\pi^2)^{1/2} N^{1/2}\hbar\omega~$.
Taking $N=10^4~ \hbox{Rb}^{87}$ atoms, this 
corresponds to $T_c^0\simeq 48$ nK. 
We must have $\hbar\omega_z\gg k_B T_c^0$ for the 
device to be effectively two-dimensional. Assuming 
$\omega_z\simeq 10^3~ \omega$, and the scattering length $a$ for Rb$^{87}$ 
atoms to be 5.8 nm\cite{Boesten}, the estimate of the interaction 
strength $g$ from Eq.~(\ref{small}) is about $0.05$. Thus the model 
Hamiltonian (\ref{tr1}) is realizable experimentally for small $g$, 
although we shall also present the results for $g=1$ to see how the 
quantum statistics is altered. We emphasize that the two-dimensional 
delta function interaction in Eq.~(\ref{tr1}) is a dimensionally reduced   
form of the three-dimensional pseudo-potential, and is not to be regarded 
as a pseudo-potential derived from two-dimensional scattering theory.
Later, we shall need to take the thermodynamic limit of this quasi 
two-dimensional system to investigate the question of strict phase 
transition. This is done by keeping $\omega _z $ at a {\it fixed}
appropriate value and letting $N\rightarrow  \infty, \omega \rightarrow 0 $,
keeping $N^{1\over 2} \omega $ constant.
Note, however, that in the alternative scheme of keeping the ratio 
$\omega _z/\omega $ fixed at a large value, and then taking the 
thermodynamic limit, would result in a three-dimensional geometry.
This is because the condition $\hbar \omega _z \gg k_B T_c^0$ would 
not be satisfied.
We now proceed to do a Thomas-Fermi calculation for this two-dimensional 
system.
\section{Thomas-Fermi Calculation}
In BEC, it is customary to write the density of bosons to be $n({\bf r})=
n_0({\bf r})+n_T({\bf r})$, where $n_0$ is the condensate density, and 
$n_T$ the density of particles occupying states other than the ground state
~\cite{Dal99}. Consider temperatures above the critical $T_c$ of the 
interacting system. Then there is no condensate, 
and $n({\bf r})=n_T({\bf r})$. 
In this situation, the one-body 
potential generated by the above zero-range interaction is 
$U(n({\bf r}))={2\pi\hbar^2\over m }g n({\bf r})$.  
Note that for $T>T_c$, the 
more sophisticated Popov approximation~\cite{Giorgini} also  
reduces to the TF approximation: 
\beq
n({\bf r})=
\int {d^2p/{(2\pi\hbar)^2}\over 
{\left[\exp [({p^2\over {2 m}}+{1\over 2} m\omega^2 r^2+
{2\pi\hbar^2\over m} g n({\bf r})-\mu)\beta]-1\right]}}~.
\label{cup}
\eeq
After performing the $p$-integration 
analytically, the number-density of particles 
($\beta=1/(k_B T)$) is given by 
\beq
Z e^{-\beta m\omega^2 r^2/2}=2 e^{\alpha u}\sinh {u\over 2}~,
\label{m1}
\eeq
where we have put $u({\bf r})=2\pi\hbar^2 n({\bf r})\beta/m$, the fugacity 
$Z=\exp (\beta\mu)$, and $\alpha=(g-1/2)$. The number of atoms in the trap 
is given by  
\beq
N={m\over {2\pi\hbar^2\beta}}\int u({\bf r}) d^2r~.
\label{mac1}
\eeq 
To obtain $n({\bf r})$ for a fixed $N$, Eqs.~(\ref{m1}) and (\ref{mac1}) 
have to be solved self-consistently. For $g=0$, these 
cannot be satisfied for $T < T_c^0 $, and the occupancy of the lowest 
quantum state (in this approximation at zero energy) 
has to be taken into account explicitly. This means that the condensate 
density $n_0$ occupying the lowest level is macroscopic below $T<T_c^0$. 
In Fig.~1, we show how this results in a discontinuity in $\mu$ at $T_c^0$ 
for $g=0$. (In numerical work throughout, we 
use arbitrary units [a.u.] with $\hbar = m = 1$   
and set $N=\omega =1$.)   
\PostScript{8}{0}{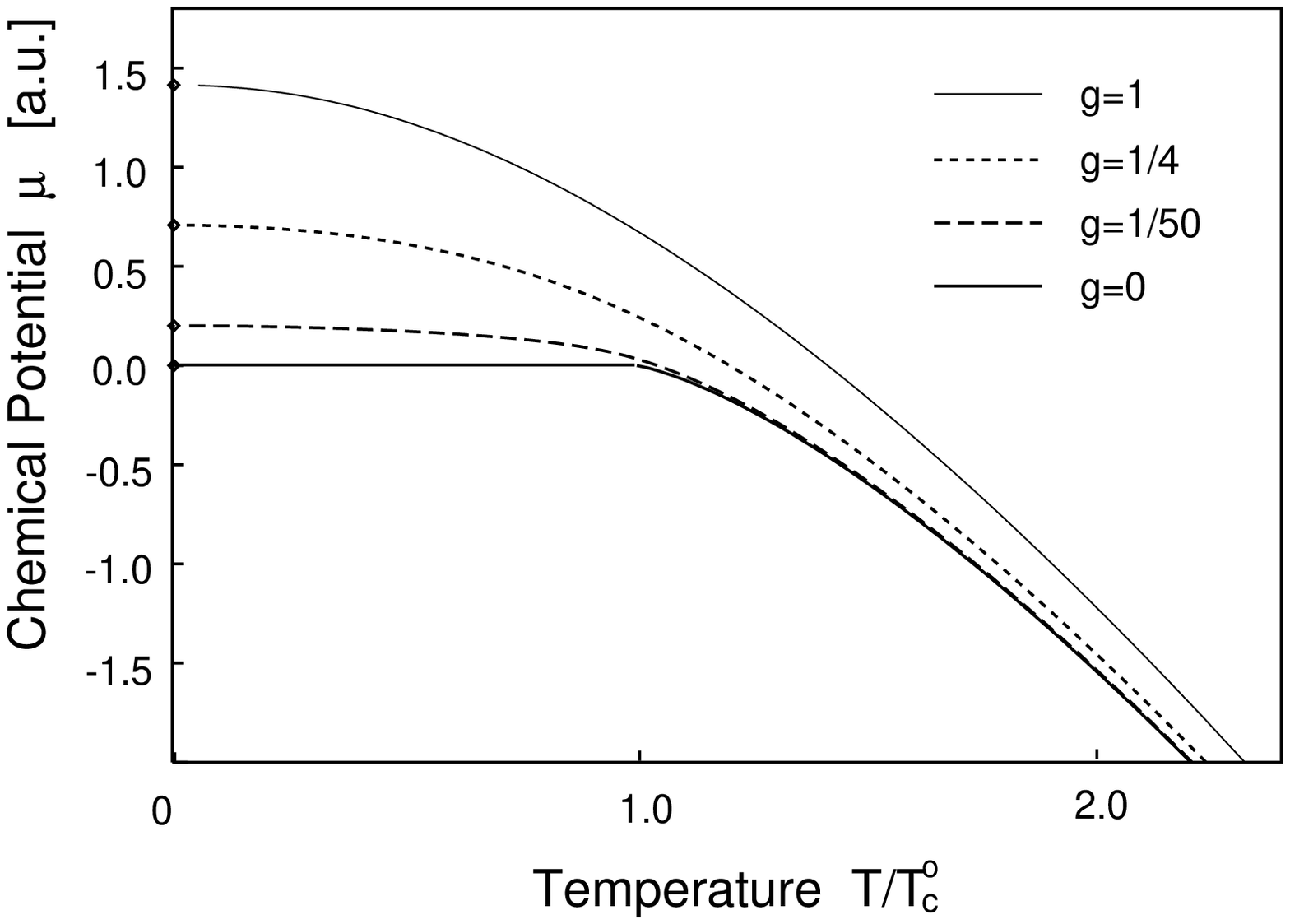}{0.5}{14}{
Chemical potential $\mu $ versus temperature $T/T_c^0$.
Self-consistent solutions of the temperature-dependent TF equations 
(\ref{m1},\ref{mac1}) are obtained for various values of the 
interaction strength $g$}{fig1}
For a nonzero positive $g$, Eqs.~(\ref{m1}) and (\ref{mac1}) can  be 
self-consistently solved right down to $T=0$ (see Fig.~1). This contradicts 
Mullin's claim~\cite{Mull98a} that there is no self-consistent solution 
of the TF equations below a nonzero $T_c$. Note that the $T\rightarrow 0$ 
limit of Eq.~(\ref{m1}) yields a nonzero spatial density 
only within the classical turning point $r_0$: 
\beq
n_0(r)={m\over {2\pi \hbar^2 g}}\left(\mu-{1\over 2}m\omega^2 r^2 \right),
~~~~r\leq r_0~,
\label{Tr4}
\eeq 
where $r_0=\sqrt{2\mu\over {m \omega^2}}$.
This gives $ N={\mu^2/{2 g (\hbar\omega)^2}}$~,
so that the chemical potential at zero temperature is given by 
$\mu(T=0)=\sqrt{2 g}~ N^{1/2}\hbar \omega~$.
Thus, the self-consistent solution  has led to the  
$T\rightarrow 0$ TF result (Eq.~(\ref{Tr4})) for $n_0$ that one would 
obtain starting from the Gross-Pitaevskii density functional~\cite{Dal99}. 
In passing we note 
that the zero-temperature density for a spin-less free Fermi gas in a  
two-dimensional harmonic potential is also given by Eq.~(\ref{Tr4}), with the 
proviso that 
the occupancy of each quantum state is $1/g$ instead of unity. Indeed, even 
though it is not experimentally feasible to extrapolate $g$ to unity, it is 
amusing to note that putting $g=1$ in Eq.~(\ref{m1}) would yield exactly the 
finite temperature {\it noninteracting} fermionic result 
\beq
n(r)={m\over {2\pi \hbar^2 \beta}} 
\ln \left(1+\exp (\mu-{1\over 2}m\omega^2  r^2)\beta \right)~\label{ferm}~.
\eeq   
These results are not accidental, and their significance (specially for 
small $g$) will presently be discussed. It is also worth noting that the 
plots in our Fig.~1 are very similar to those of Fig.~2 for a noninteracting 
Haldane gas in Ref.~\cite{Sen95}.                            

It is important to note that the situation is very different for an 
interacting gas in a three-dimensional trap, where the TF equation 
analogous to Eq.~(\ref{cup}) has no 
self-consistent solution for $T<T_c$. One then has to solve the Bogoliubov 
equations self-consistently, say in the Popov approximation~\cite{Dal99}. For 
$T>T_c$, this procedure reduces to solving the  
temperature-dependent TF equation. Qualitatively, the calculated curves of 
$\mu$ vs $T$ for the three-dimensional trap (shown in Fig.~26 
of \cite{Dal99}) look similar to our Fig.~1. This is somewhat deceptive, 
since the curves in our Fig.~1 are obtained in the TF approximation for all 
$T$, whereas the Popov approximation is used for $T<T_c$ in the 
three-dimensional case. The behavior of the so-called 
``release energy'', which is the total minus the harmonic oscillator 
energy, has a very interesting structure below and at $T_c$ in the 
three-dimensional case. 
\PostScript{8.5}{0}{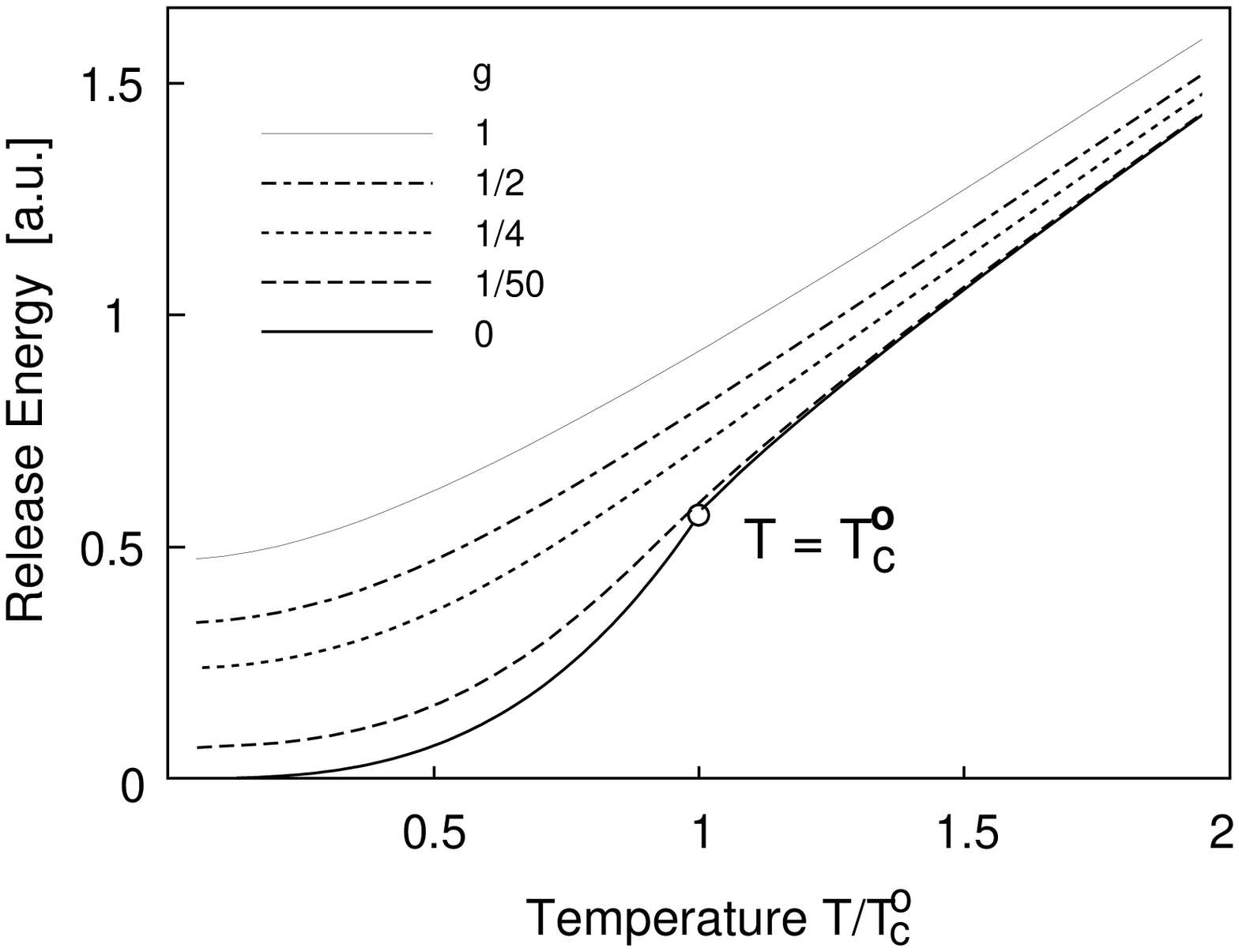}{0.5}{14}{Release energy versus 
temperature $T/T_c^0$ from the self-consistent solution of the 
temperature-dependent TF equation. For comparison, the 
noninteracting $g=0$ case is also shown}{fig2}
\noindent
This is also confirmed experimentally by Ensher {\it et al.}~\cite{Ensher} 
for Rb atoms, who find a discontinuous behavior of the deduced specific heat. 
By contrast, as is shown in Fig.~2 for the two-dimensional problem, 
{\it no such structure is found for nonzero $g$
in the calculated release energy}.   
Experimentally, in three-dimensional traps, the appearance of a 
condensate is manifested as a sharp peak in the spatial density profile of 
the gas~\cite{And95}. 
For the interacting two-dimensional gas, as shown in Fig.~3,
the central density rises gradually rather than abruptly
with the lowering of temperature.
(This, however, may be a limitation of the TF model, see~\cite{Kett96}.) 
\PostScript{8}{0}{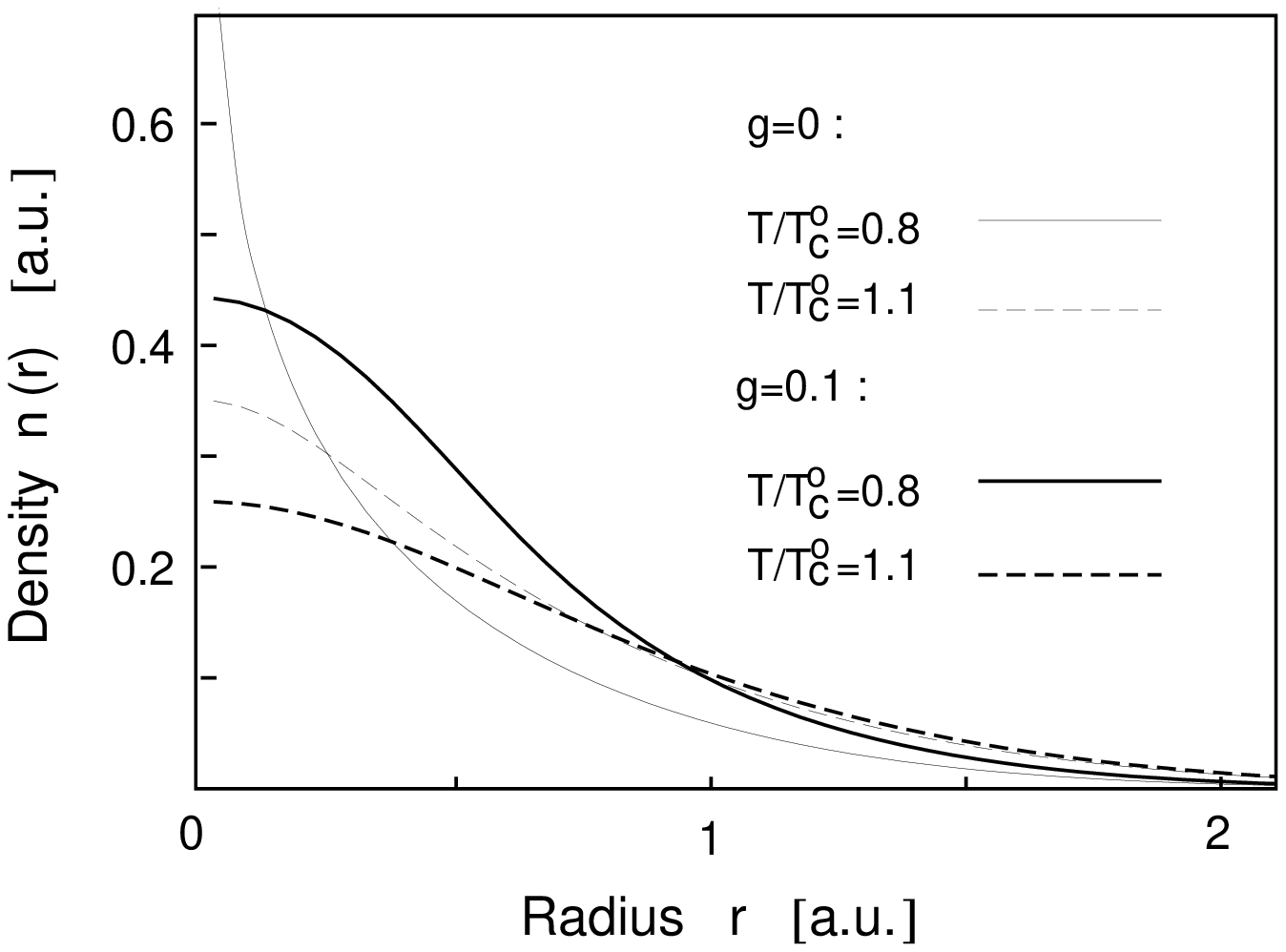}{0.5}{14}{
Self-consistent TF density distribution, with unit normalization, 
as a function of the radial distance $r$ for temperatures above and 
below $T_c^0$. The noninteracting $g=0$ case is shown for comparison}{fig3}
We now show analytically, starting from Eqs.~(\ref{m1},\ref{mac1}),  
that for $g>0$ there can be no phase transition. Formally, these equations  
bear a striking similarity with Eqs. (29,30) of Sutherland's classic 
paper~\cite{Suth71} on the thermodynamics of the Calogero-Sutherland model 
(CSM)~\cite{Calo69}. We exploit this for our proof, going to the 
thermodynamic limit $N\rightarrow \infty$, $\omega \rightarrow 0$, with 
$N^{1/2}\omega =1$. In this limit, the $r$-dependence in Eq.~(\ref{m1}) drops 
out, and we get 
\beq
Z=\left[\exp(g u)-\exp((g-1)u)\right]~.
\label{mac}
\eeq
From this, we see that for $g=0$, $u=-\ln(1-Z)$, so $u$ has a branch-point 
at $Z=1$, i.e., at $\mu=0$. This is where BEC takes place for the ideal 
Bose gas. For $g=1$, on the other hand, Eq.~(\ref{mac}) gives 
$u=\ln(1+Z)$, with a branch point at $Z=-1$. Similarly, for $g=1/2$, it is 
easily checked that $u$ has branch points at $Z=\pm 2i$. More generally,  
the location of the branch points of $u$ (which is proportional to the 
spatial density) in the complex $Z$-plane as a function 
of $g$ are shown in Fig.~4, which is identical to Sutherland's Fig.~1 for 
CSM in~\cite{Suth71}. The only singularity on the positive real axis is 
at $g=0$, and for $g>0$ there is no phase transition.                        
\PostScript{7}{0}{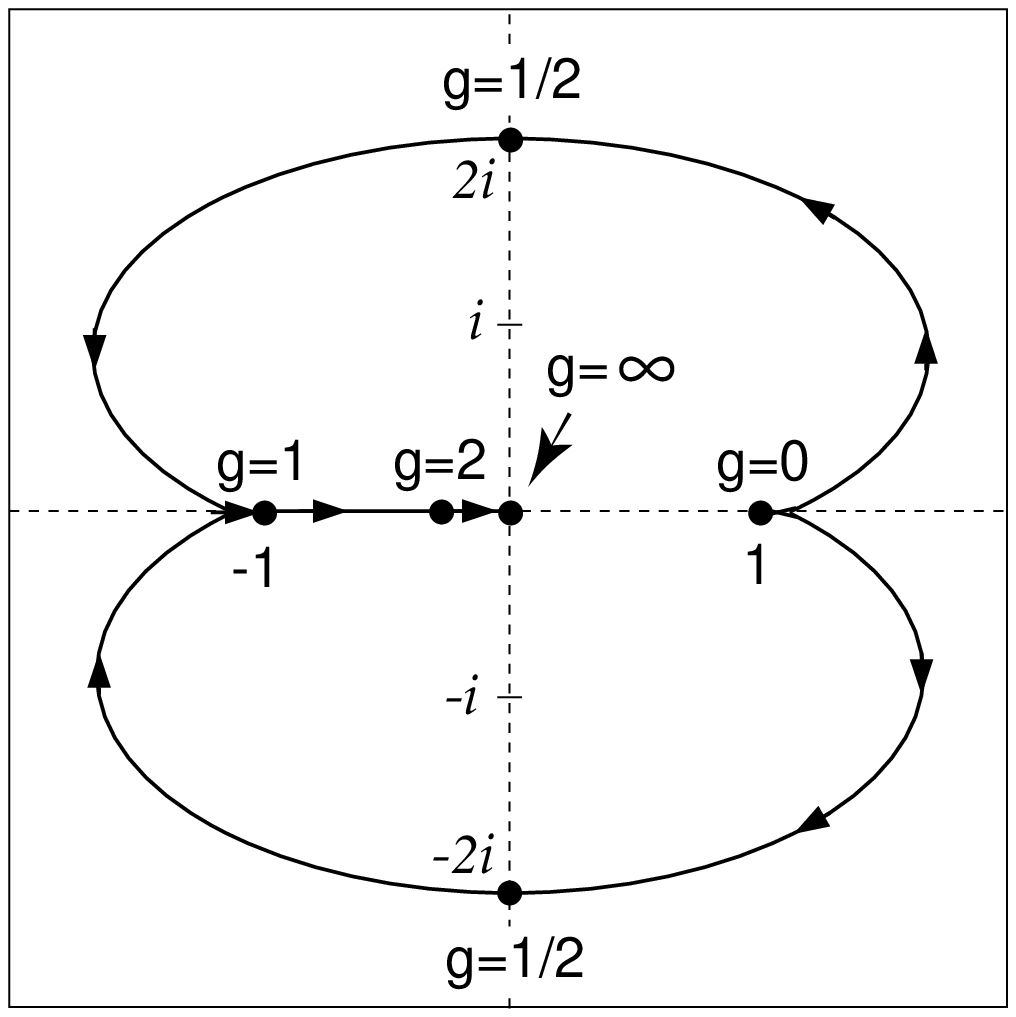}{0.5}{14}
{Location of the branch points of $u$ in the complex $Z$-plane as a function 
of $g$ (cf.~\cite{Suth71}) }{fig4}

\section{Equivalence to fractional exclusion statistics}

The two-dimensional delta function interaction 
considered in Eq.~(\ref{tr1}) is very special in the sense that it is 
scale-independent~\cite{Pit97}. 
A two-body potential that varies as the inverse square 
of the distance between the two particles also shares this property. In one 
dimension, the latter is the much-studied exactly solvable $N$-particle 
CSM, which is known to obey~\cite{MSI,Ha,Isakov} 
Haldane's fractional exclusion 
statistics (FES). By this we 
mean that either one can solve the $N$-body interacting bosonic or fermionic 
problem with the pair-wise inverse square interaction, or map it onto a 
noninteracting set of $N$-particles obeying FES. 
The statistical occupancy factor for an ideal FES gas is given by
~\cite{Isakov,Wu} 
$\eta_i=(w_i+g)^{-1}$, where 
\beq
w_i^g (1+w_i)^{1-g}=exp[(\epsilon_i-\mu)\beta].
\label{fesw} 
\eeq
We note that FES is characterized by a ``statistical factor'' $g$, which will 
presently be identified with the dimensionless interaction strength of 
the delta function interaction in Eq.~(\ref{tr1}). 
The occupancy factor in FES reduces to the usual bosonic 
or the fermionic one for $g=0$ or $1$, respectively.

In order to demonstrate explicitly the equivalence between our 
two-dimensional  interacting boson model and FES, let us consider 
the analogue of Eq.~(\ref{cup}) for the number density of a 
gas of non-interacting ``haldons'',
\beq
n({\bf r}) = \int \frac{d^2p/(2\pi\hbar)^2}{w+g},
\eeq
where $w$ depends on $p$ via Eq.~(\ref{fesw}). The number density
can be rewritten as an integral over $w$ by noting that Eq.~(\ref{fesw})
gives
$[g/w + (1-g)/(1+w)]dw = (\beta/2\pi m) d^2p$, so that
\beq
n({\bf r}) = \frac{m}{2\pi\hbar^2\beta} \int_{w_0}^{\infty} \frac{dw}{w(1+w)}
           = \frac{m}{2\pi\hbar^2\beta} \ln\frac{1+w_0}{w_0},
\label{X}             
\eeq
with $w_0 \equiv w(p=0)$ given by
\beq
w_0^g(1+w_0)^{1-g} = Z^{-1} e^{\beta m\omega^2r^2/2}.
\label{XX}
\eeq
Eliminating $w_0$ from Eqs.~(\ref{X}) and (\ref{XX}) just gives Eq.~(\ref{m1}),
which shows that the number densities of the two models are the same. 
We proceed to prove that the same is the case for the energy density,
which for the FES gas is given by
\beq
E({\bf r}) = \int \frac{d^2p/(2\pi\hbar)^2}{w+g}
             \left[ \frac{p^2}{2m} + \frac{m\omega^2}{2}r^2 \right].
\eeq
Changing variables from $p$ to $w$ as above and using Eq.~(\ref{fesw})
to substitute
\beq
\beta [p^2/2m + m\omega^2 r^2/2] = \beta\mu+g\ln w + (1-g)\ln (1+w),
\eeq
the energy density takes the form
\beq
E({\bf r})= \frac{m}{2\pi\hbar^2\beta^2} \int_{w_0}^{\infty}
   \frac{dw}{w(1+w)} \left[ g\ln\frac{w}{1+w} + \ln (1+w)+\beta\mu \right].
\label{eint}
\eeq
The first term of the integrand is a total derivative, giving a
contribution
\beq 
(-mg/4\pi\hbar^2\beta^2)\left( \ln \frac{w_0}{1+w_0}\right)^2 
  = (-\pi\hbar^2g/m) (n({\bf r}))^2
  = -\frac{g}{2\beta^2} \int_{w_0}^{\infty} \frac{dw}{w(1+w)} n({\bf r}) 
\label{cont}
\eeq
to the energy density. 
With the variable substitution
$w = w_0 + (1+w_0)(\exp(y)-1)$, the remainder of the integral (\ref{eint})
takes the form
\beq
E({\bf r}) + \frac{\pi\hbar^2 g}{m}\left( n({\bf r}) \right)^2 =
 \frac{m}{2\pi\hbar^2\beta^2} \int_0^{\infty}
  \frac{dy}{(1+w_0)e^y - 1} \left[ y+\ln(1+w_0)+\beta\mu \right].
\label{preint}
\eeq
Here, $w_0$ can be eliminated by using the relation
\beq
1+w_0 = \exp \left\{ \beta\left[ m\omega^2r^2/2 
+ (2\pi\hbar^2g/m) n({\bf r}) -\mu \right] \right\},
\label{w0}
\eeq
which is found by combining Eqs.~(\ref{X}) and (\ref{m1}). Further
identifying $y=\beta p^2/2m$ and collecting all terms, we find the
final form of the energy density integral,
\beq
E({\bf r}) = \int {d^2p/{(2\pi\hbar)^2}
 \left[{p^2\over {2 m}}+{1\over 2} m\omega^2 r^2
+ {\pi\hbar^2\over m} g n({\bf r})\right]
\over 
{\left[\exp [({p^2\over {2 m}}+{1\over 2} m\omega^2 r^2+
{2\pi\hbar^2\over m} g n({\bf r})-\mu)\beta]-1\right]}}.
\label{beint}
\eeq
But this is just the energy density of the interacting boson gas.
Note that Eq.~(\ref{w0}) brings in a term $ \sim g n({\bf r}) $ on the
right-hand side of Eq.~(\ref{preint}), which combines with the
one in Eq.~(\ref{cont}) to give the correct coefficient
$\pi\hbar^2 g/m$ in Eq.~(\ref{beint}).
Thus, we have demonstrated that 
{\it the interacting bosons may be mapped on to 
noninteracting particles obeying FES}. Although above we have assumed a 
harmonic trap, note that our proof holds for any 
shape of the confining trap potential.

For a dilute gas in the thermodynamic limit obeying FES in a $D-$dimensional 
system, the statistical factor $g$ may be related~\cite{MSII,Isakov2} to the 
{\it high-temperature limit} of the second virial coefficient 
$B_2$~\cite{Ukirk}:
\beq
g-{1\over 2}=2^{D/2} B_2~.
\label{ron3}
\eeq
For the parameter $g$ from Eq.~(\ref{ron3}) to be meaningful, it should be 
temperature-independent. This is possible if the interaction potential 
is scale-independent. This requirement is satisfied by an inverse-square 
potential in any dimension. In one dimension, the corresponding potential 
is the CSM mentioned before. Interestingly, the two-dimensional case is very 
special, since in this case the area-integral of 
any potential, when it exists, turns out to be proportional to the 
statistical parameter. To see this, consider, for $(D=2)$, the 
$\beta\rightarrow 0$ limit of $B_2$ given in \cite{Ukirk}: 
\beq
B_2={m\over {4\pi\hbar^2}} M_0~\pm {1\over 4}~,
\eeq 
where $M_0=\int~d^2r V(\bf{r})$. We have assumed that the potential $V$ 
is well-behaved, so that 
the expansion $\exp(-\beta V)=[1-\beta V + (\beta V)^2/2 -...]$ is valid.
Terms containing $\int~d^2r V^n$ for $n\ge 2$ do not contribute in the 
high-temperature limit, since $\beta^{(n-1)}\rightarrow 0$. Considering bosons 
in $D=2$, Eq.~(\ref{ron3}) gives 
$M_0=(2\pi\hbar^2/m) g~$.
The zero-range {\it  effective} potential in Eq.~(\ref{tr1}) has 
precisely this 
moment $M_0$, and its strength $g$ is thus the statistical parameter 
for Haldane statistics.

\section{DISCUSSION}
In summary, we have shown, using the Thomas-Fermi method, that the effect of 
a repulsive zero-range interaction between 
the atoms in the ``two-dimensional'' trap is more drastic than 
the corresponding three-dimensional case.
Even in the thermodynamic limit, there is no strict phase transition 
with a repulsive zero-range two-body interaction, as shown by the behavior 
of the spatial density in the complex $Z$-plane.
We have also demonstrated that such interacting bosons in a two 
dimensional trap have identical bulk thermodynamic properties as those of 
noninteracting ``haldons'' in the same trap, obeying FES. 

The above conclusions have been reached using a two-dimensional zero-range 
interaction with a fixed coupling constant, which was obtained by a  
dimensional reduction of the three-dimensional pseudo-potential, as explained 
in the introduction. 
It should be noted, however, that in a {\it strict} two-dimensional problem, 
the pseudo-potential is more complicated. As emphasized by 
Schick~\cite{schick}, 
the scattering cross section for binary collision of two bosons due to a 
hard-disc potential of radius $a$ diverges in the low-energy limit :
\beq
\sigma \rightarrow {\pi^2\over {k(\ln ka)^2}}~,~~~~(ka)<<1~,
\label{kdep}
\eeq  
where $k$ is the relative wave number. This is a very different behaviour 
from the three-dimensional problem, 
where the total cross section goes to a constant value proportional to 
$a^2$. Thus, in strictly two-dimensions, the strength of the pseudo-potential 
should be $k$-dependent. 

Another way to define a pseudo potential for a 
very dilute gas would be to have a zero-range form with a 
density-dependent strength such that in Born approximation it reproduces the 
ground-state energy per particle to the lowest order. Schick had shown 
that this is given by $E/N=-2\pi n (\ln na^2)^{-1}$, where $n$ is the 
average number density of the bosons. Shevchenko~\cite{shev}, 
in a detailed analysis of bosons in a trap, thus adopted a zero-range 
potential with the $g$ of our Eq.(\ref{small}) replaced by the 
density-dependent factor $\ln^{-1}(1/(n a^2))$. He came to the 
conclusion that whereas this  repulsive interaction prevents strict BEC, 
superfluidity sets in~\cite{ber} at a temperature close to the critical value 
$T_c^0=(6/\pi^2)^{1/2} N^{1/2}\hbar\omega$ of the ideal gas. This transition 
to superfluidity for the compressed fraction of the gas near the center of 
the trap takes place through the Berezinskii-Kosterlitz-Thouless~(BKT)
~\cite{bkt} phenomenon of bound vortex pairs, much as in the translationally 
invariant two-dimensional interacting gas.
  
In our simple Thomas-Fermi treatment of the two-dimensional problem with 
a zero-range pseudo potential of constant coupling strength, we cannot  
examine the fluctuations in the phase of the ``quasi-condensate''
~\cite{kagan}, 
nor can we comment on the formation of BKT-vortices. 
Despite these limitations, we 
find, for our model, the transformation of Bose to Haldane statistics 
through the zero-range interaction. This, arguably, is the most interesting 
point of our paper.

\smallskip

Two of the authors (R.K.B. and S.M.R.) would like to thank $E.C.T.^*$ at 
Trento where much of the work was initiated during a workshop. In this 
connection, we benefited greatly from many discussions with Ben Mottelson. 
We are also indebted to M.V.N. Murthy for helping to shape many of the ideas 
in this paper, and to T. H. Hansson for help in the derivation in 
the third section. A.G.C. would like to thank the department of Physics and 
Astronomy of McMaster university for hospitality. This research was supported 
by Natural Sciences and Engineering Research Council of Canada
and the TMR programme of the European 
Community under contract ERBFMBICT972405.



\begin{thebibliography}{99}

\narrowtext

\bibitem{Bag91} V. Bagnato and D. Kleppner, Phys. Rev. A {\bf 44}, 
7439 (1991).

\bibitem{Kett96}
W. Ketterle and N. J. van Druten, Phys. Rev. A {\bf 54}, 656 (1996).

\bibitem{Gross96}
S. Grossman and M. H. Holthaus, Phys. Rev. E {\bf 54}, 3495 (1996).

\bibitem{Haug98}
T. Haugset and H. Haugerud, Phys. Rev. A {\bf 57}, 3809 (1998).

\bibitem{Dal99} F. Dalfovo {\it et al.}, Rev. Mod. Phys. {\bf 71}, 463 (1999).

\bibitem{Haldane91} F. D. M. Haldane, Phys. Rev. Lett. {\bf 67}, 937 (1991).

\bibitem{Fermi36} E. Fermi, Ricerca Sci. {\bf 7}, 13 (1936).
\bibitem{Huang89} K. Huang and C. N. Yang, Phys. Rev. {\bf 105}, 767 (1957); 
K. Huang, Int. J. Mod. Phys. A {\bf 4}, 1037 (1989).

\bibitem{Pit97} L. P. Pitaevskii and A. Rosch, Phys. Rev. A {\bf 55}, 
R853 (1997).

\bibitem{Ber98} T.~Papenbrock and G.~F.~Bertsch, Phys. Rev. A {\bf 58}, 4854 (1998)

\bibitem{Mull97b} W. J. Mullin, J. Low Temp. Phys. {\bf 106}, 615 (1997).

\bibitem{Mull98a} W. J. Mullin, J. Low Temp. Phys. {\bf 110}, 167 (1998).

\bibitem{Boesten} H. M. J. Boesten {\it et al}., Phys. Rev. {\bf A55}, 
636 (1997).

\bibitem{Giorgini} S. Giorgini, L. P. Pitaevskii, and S. Stringari, 
Phys. Rev. {\bf A 54} R4633 (1996). Note that Eqs.(6-11) of this paper  
reduce to the TF approximation when the condensate density $n_0$ is 
set to zero.

\bibitem{Sen95} Diptiman Sen and R. K. Bhaduri, Phys.~Rev.~Lett. {\bf 74}, 
1395 (1995). In the thermodynamic limit, given the same spatial density $n$, 
our Eq.~(\ref{cup}) reduces exactly to Eq.~(26) of this reference.  

\bibitem{Ensher} J. R. Ensher {\it et al.},
Phys. Rev. Lett. {\bf 77}, 4984 (1996).

\bibitem{And95} M. H. Anderson {\it et al.}, Science {\bf 269}, 198 (1995).

\bibitem{Suth71} B. Sutherland, J. Math. Phys. {\bf 12}, 251 (1971).

\bibitem{Calo69} F. Calogero, J. Math. Phys. {\bf 10}, 2197 (1969); 
B. Sutherland, J. Math. Phys. {\bf 12}, 246 (1971); Phys. Rev. A {\bf 4}, 
2019 (1971); Phys. Rev. A {\bf 5}, 1372 (1972).

\bibitem{MSI} M. V. N. Murthy and R. Shankar, Phys. Rev. Lett. 
{\bf 73}, 3331(1994).

\bibitem{Ha} Z. N. C. Ha, Phys. Rev. Lett. {\bf 73}, 1574 (1994); 
Nucl. Phys. B{\bf 435}, 604 (1995).  

\bibitem{Isakov} S. B. Isakov, Phys. Rev. Lett. {\bf 73}, 2150 (1994); 
D. Bernard and Yong Shi Wu, UU-HEP/9403, cond-mat/9404015.  

\bibitem{Wu} Yong Shi Wu, Phys. Rev. Lett. {\bf 73}, 922 (1994). 

\bibitem{MSII} M. V. N. Murthy and R. Shankar, Phys. Rev. Lett. 
{\bf 72}, 3629 (1994).

\bibitem{Isakov2} S. B. Isakov {\it et al.}, Phys. Lett. 
A {\bf 212}, 299 (1996).

\bibitem{Ukirk} G. E. Uhlenbeck and L. Gropper, Phys. Rev. {\bf 41}, 79 
(1932); J. G. Kirkwood, Phys. Rev. {\bf 44}, 31 (1933). The semiclassical 
expression for the second virial coefficient is given by 
$B_2 ={(\lambda)^{D/2}\over 2} 
\int dr^D~ \left[1-\exp(-\beta V(\bf r))\right]~\pm  2^{-(1+D/2)}$~.
The positive (negative) sign in the last term is for fermions (bosons), and  
$\lambda= ({m/{2\pi\hbar^2\beta}})$.

\bibitem{schick} M. Schick, Phys. Rev. {\bf A3}, 1067 (1971).

\bibitem{shev} S. I. Scevchenko, Sov. Phys. JETP {\bf 73}, 1010 (1991).

\bibitem{ber} V. L. Berezinskii, Sov. Phys. JETP {\bf 32}, 493 (1971); 
{\bf 34}, 610 (1972). 

\bibitem{bkt} M. Kosterlitz and D. J. Thouless, J. Phys. {\bf C6}, 1181 
(1973); M. Kosterlitz, J. Phys. {\bf C7}, 1046 (1974).

\bibitem{kagan} Yu. Kagan, B. V. Svistunov and G. V. Shlyapnikov, Sov. Phys. 
JETP {\bf 66}, 480 (1987).

\end{thebibliography}
\end{document}